# Heavy Ion Induced SEU Sensitivity Evaluation of 3D Integrated SRAMs[*]

Xuebing Cao(曹雪兵)[1], Liyi Xiao(肖立伊)[1,2;1)], Mingxue Huo(霍明学)[3], Tianqi Wang(王天琦)[3], Anlong Li(李安龙)[1], Chunhua Qi(齐春华)[1], Jinxiang Wang(王进祥)[1]

[1]Microelectronics Center, Harbin Institute of Technology, Harbin, 150001, China
[2]Harbin Institute of Technology Shenzhen Graduate School, Shenzhen, 518055, China
[3]Research Center of Basic Space Science, Harbin Institute of Technology, Harbin, 150001, China

**Abstract**：Heavy ions induced single event upset (SEU) sensitivity of three-dimensional (3D) integrated SRAMs are evaluated by using Monte Carlo simulation methods based on Geant4. The cross sections of SEUs and Multi Cell Upsets (MCUs) for 3D SRAM are simulated by using heavy ions with different energies and LETs. The results show that the sensitivity of different die of 3D SRAM has obvious discrepancies at low LET. Average percentage of MCUs of 3D SRAMs rises from 17.2% to 32.95% when LET increases from 42.19 MeV•cm$^2$/mg to 58.57 MeV•cm$^2$/mg. As for a certain LET, the percentage of MCUs shows a notable distinction between face-to-face structure and back-to-face structure. For back-to-face structure, the percentage of MCUs increases with the deeper die. However, the face-to-face die presents the relatively low percentage of MCUs. The comparison of SEU cross sections for planar SRAMs and experiment data are conducted to indicate the effectiveness of our simulation method. Finally, we compare the upset cross sections of planar process and 3D integrated SRAMs. Results demonstrate that the sensitivity of 3D SRAMs is not more than that of planar SRAMs and the 3D structure can become a great potential application for aerospace and military domain.

## 1. Introduction

As a result of technology scaling, single event effects (SEEs) due to high-energy particles striking have been a main challenge for modern nanoscale CMOS ICs. Single event upsets (SEUs) result from particles induced storage state change in sequential logic circuits are still dominate in 32nm technology process [1]. In particular, as the decreasing critical charge and distance between transistors, the single particle can induce increasing multiple cell upsets (MCUs) due to the deposited charges are collected by adjacent cells in SRAMs. The various particles, for instance, α particle emitted from the radioactive impurities in package materials and neutron or proton come from high-energy cosmic rays [2,3]. Specially, the negligible low energy proton and muon in the past have become significant considering in single event evaluation [4,5]. Comparing with aforementioned particles, the heavy ions deposit more energy per unit depth along its track by directly localize ionization or indirectly nuclear reaction. Consequently, heavy ions have always been an important single event source. The evaluation of sensitivity result from heavy ions are considered necessary [6].

Due to allowing the complexity and density of the circuits, 3DIC technology as an alternative technology for relieving the negative effects of technology scales has aroused great attention in the field of design and research of integrated circuit. 3DIC technology has changed the traditional integration way by stacking two or more layers of transistors vertically and horizontally [7]. By bonding two or more IC wafers, 3DIC technologies can increase the density of transistors per unit area and the flexibility of placement and routing with the aid of TSV (through silicon vias) technique. Furthermore,

[*] This work was supported by the Fundamental Research Funds for the Central Universities (Grant No.HIT.KISTP.201404), Harbin science and innovation research special fund (2015RAXXJ003), and Special found for development of Shenzhen strategic emerging industries (JCYJ20150625142543456).

1) corresponding author: Liyi Xiao; phone: 86-045186413405-804; fax: 86-045186413405-804; e-mail: xiaoly@hit.edu.cn

transportation delay time between different logical units that located in different dies can been significantly decreased by optimization of IC's layout. For hybrid IC designer, 3DIC technology allows an optimization of the design technology for specific function or particular application, electromagnetic interference can be reduced significantly by placing digital and analog module in different die and the noise performance of circuit systems can be improved.

The advantages of the 3DIC integration technology have an attractive potentiality for space and military applications due to the consideration in limitation of volume and performance of spacecraft. So it is prerequisite to research the vulnerability of 3DIC which operated in a complicated environment. The sensitivity of planar SRAMs induced by heavy ions have been researched widely. Robert et al. studied the impact of individual ionizing heavy ions on SRAMs as early as 1997 [8]. Damien et al. analyzed the MBU cross-sections of heavy ions for 90nm SRAMs [9]. Boorboor et al. investigated the impact of radial dose effect on SEUs due to heavy ions radiation [10]. However, the study on SEE characteristic of 3DIC to various particles, especially for heavy ions, are still not very explicit. Pascale et al. reported the experiment results of 3×64k 150nm SOI SRAM, proton irradiation and 14MeV neutron irradiation results are discussed. The results dedicated that upset cross sections for vertically integrated SRAMs are similar for all three dies, but for heavy ions, no results were presented [11]. Peng Li et al. researched the impact of heavy ion species and energy on SEE characteristics of 3DIC. However, the relationship of upset cross section with LET and MCUs phenomenon not present in this work [12]. In this paper we researched the single event upset sensitivity in SRAMs that are 3D integrated in five dies. A nested sensitive volumes models based NCSU 45nm PDK are calibrated by using TCAD tool. The 3D SRAMs simulation model with five dies are built in Geant4. The heavy ions of different species and energies are simulated. Our research obtained the following achievements. 1), No obvious die-to-die effects are observed for 3D SRAMs when LET excess 15.63 MeV·cm$^2$/mg, the upset cross sections of 3D SRAMs show obvious discrepancies at low LET. 2), Effects of MCUs for 3D SRAMs are investigated, simulation results indicated that the percentage of MCUs increases obviously following the rising LET. There are significant differences in the percentage of MCUs between back-to-face structure and face-to-face structure. 3), Comparisons of cross sections of planar SRAM and experiment data, planar and 3D SRAMs are presented. The results indicated that effectiveness of evaluation methods in our simulations and the vulnerability of 3D SRAMs are not more than that of planar process, thus 3DIC technology is expected to be applied in the field of aerospace and military in the future.

## 2. Extraction of Sensitive Volumes

The basic structure of a SRAM cell indicates in Fig.1. Usually, every SRAM cell have two sensitive transistors due to its storage state for Q=0, QN=1 or Q=1, QN=0, as show in Fig.1. For example, when Q=1, QN=0, P1 and N2 are conducted, meanwhile P2 and N1 are cut off. If an ion strikes on the drain region of P2, the drain regions of off-transistors begin to collect charges due to ions striking on the space-charge region located in reverse biased p-n junction. If the collected charges exceed the critical charge, the transistor P2 will been conducted, which may force the storage state of SRAM changing to Q=0, QN=1 and a SEU will be arisen. If the generated charges are collected by sensitive transistors of adjacent SRAM cells, the storage state of those SRAMs may all changed due to this event, that refer to multiple cell upsets (MCUs).

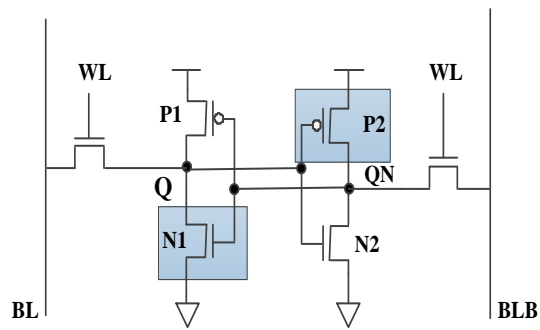

Fig.1 Basic structure of 6-T SRAM cell, P2 and N1 become sensitive transistor when Q=1 and QN=0.

## 2.1 Calibration of sensitive transistors

To analyze the influenced region due to ionization when ions strike on the sensitive transistors of SRAMs, the TCAD models are calibrated from NCSU FreePDK3D45 v1.1 firstly. The width and length of P1 and P2 is 70nm and 55nm, and the width and length of N1 and N2 are 240nm and 55nm, respectively. Transistors are built on a P-substrate that implanted a constant doping of $1 \times 10^{16}$ cm$^{-3}$. The size of P-substrate in simulation is $10 \times 10 \times 10 \ \mu m$. For all TCAD simulations in this work, the physical models are used as follows: 1) Fermi-Dirac statistics, 2) band-gap narrowing, 3) Shockley–Read–Hall (SRH) recombination and Auger recombination, 4) doping, electric field, carrier-carrier scattering, and interface scattering effects on mobility, 5) a hydrodynamic model is used for carrier transport [13]. The electrical characteristics of both PDK and TCAD models are matched, $I_{drain}$-$V_{drain}$ and $I_{drain}$-$V_{gate}$ curves of both TCAD and SPICE are shown in Fig.2 and Fig.3.

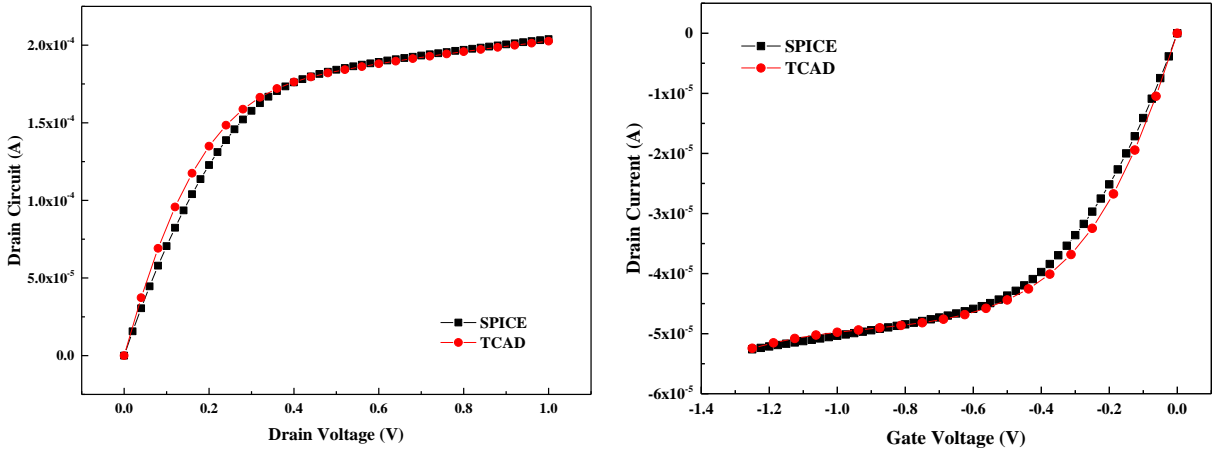

Fig.2 $I_{drain}$-$V_{drain}$ current curves of SPICE and TCAD model, where (a) is NMOS and (b) is PMOS.

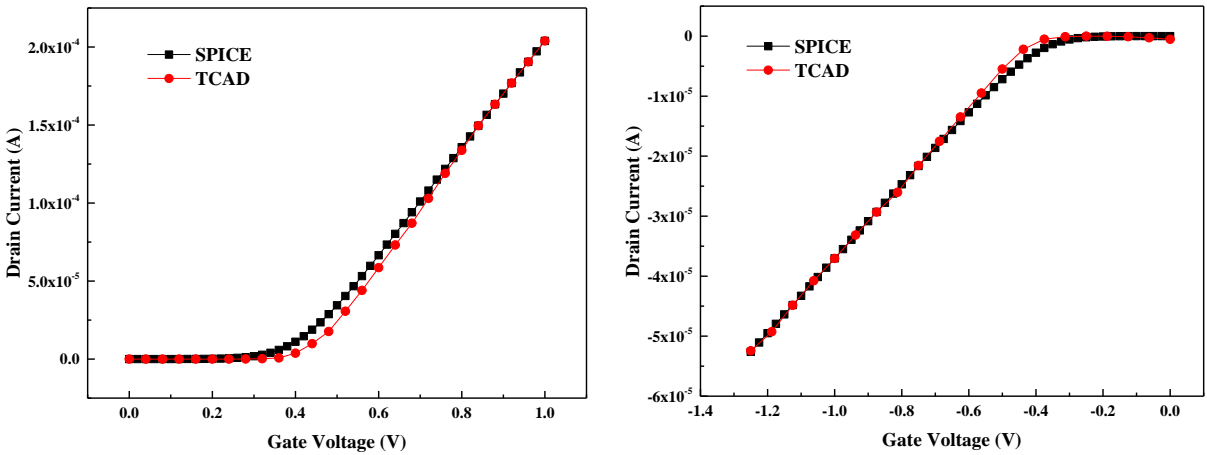

Fig.3 $I_{drain}$-$V_{gate}$ current curves of SPICE and TCAD model, where (a) is NMOS and (b) is PMOS.

## 2.2 Partition of sensitive volumes

Normally, the classic RPP model are used to simulate the sensitive regions in a transistor. However, the technology scaling has skewed the accuracy of RPP model that is used to estimate the cross sections for more advanced SRAMs [14]. In this work, the nested sensitive volume models are calibrated. The size and position are obtained by combining TCAD and SPICE simulation. The sensitive transistors are built by TCAD using the calibration results in 2.1 and the other transistors adopted SPICE models. The heavy ions with LET=5, 10 and 15 MeV-cm$^2$/mg strike on the center of the drain region of turn-off NMOS and PMOS, respectively. Heavy ion strikes are modeled as a Gaussian distribution along the striking path. Charge track radius and length is 50nm and 10μm respectively, and ions striking occur at 1ns. The position of ions striking on transistor increases at step of 0.02 $\mu$m from the center of drain to 0.2 $\mu$m, and increases at step of 0.05 $\mu$m to 0.5 $\mu$m. Fig.4 shows the collected charges at different position from the center of drain for NMOS and PMOS transistors. It can be found that the sensitivity of turn-off transistor has an obvious change within 0.2 $\mu$m. The amount of collected charge in regions that exceeding 0.2 $\mu$m decreases slightly, that means the collected charges by drain in this regions have little influence to sensitivity. So our nested SVs are limited in the range of 0.2 $\mu$m from the center of drain for sensitive transistor in a SRAM cell.

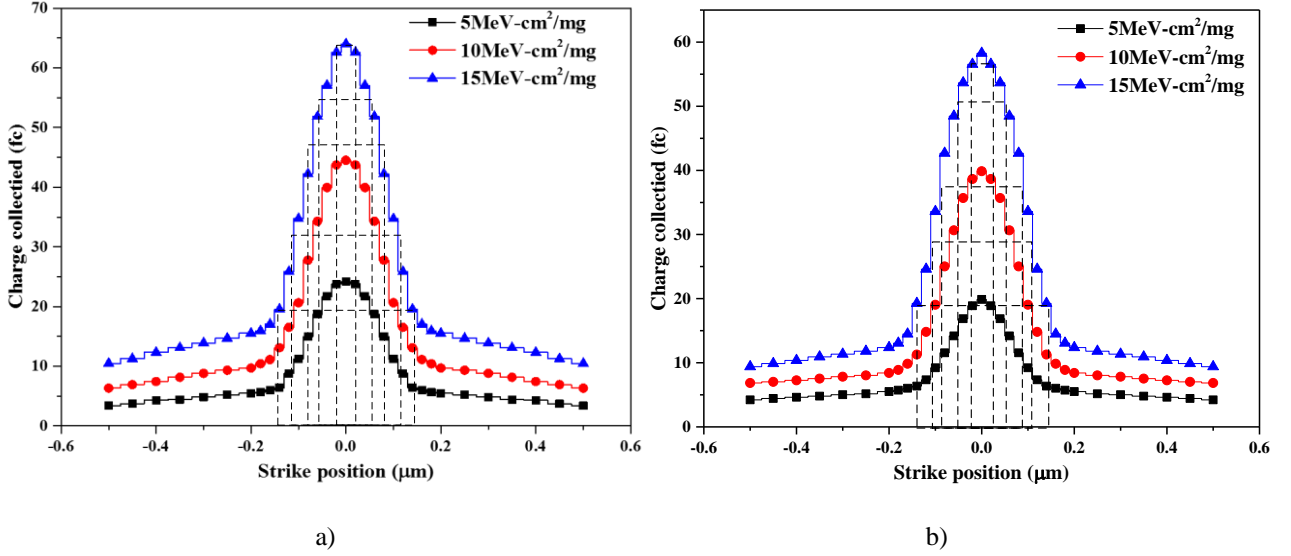

a)  b)

Fig.4 The collected charges of sensitive transistor at different position.

The depth of SVs can be estimate by the follows:

$$L_i = Q_{i,collected} / LET \tag{1}$$

Where $Q_{i,collected}$ is the amount of collected charge in the $i$th SV and LET is parameter of incident ion, it means that the energy loss per unit length when ions pass through material, normally the unit is Mev-cm$^2$/mg, it can be calculated by Pico coulomb unit (fC/$\mu$m) by multiplying by 10 [15].

The collection coefficient $\alpha_i$ of each sensitive volume can be calculated as follows:

$$\alpha_i = \frac{Q_{collect\_svi}}{Q_{collect\_drain}} \tag{2}$$

Where, $Q_{collect\_drain}$ is charge collected by drain when ions strike the center of drain region, $Q_{collect\_svi}$ is average charge collected by $i$th sensitive volume. Five sensitive volumes of turn-off PMOS and NMOS for a SRAM cell are built based on this process.

## 3. Monte Carlo Simulation

### 3.1 Model construction of 3D SRAMs

The 3D SRAMs model for Monte Carlo simulations in this paper is indicated in Fig.5. It is a refined model and designed by reference NCSU FreePDK3D45 design kit [16]. The 3D SRAMs simulation model is built of five dies; each die includes 16×16 SRAM cells. The nested five SVs of PMOS and NMOS are located in substrate of each die. The structure size of all sensitive volumes are calibrated using TCAD simulations in Section 2.2. In this 3D simulation model, first wafer is bonded by face-to-face structure, and the other wafers are designed by using back-to-face bonding. Each die has 10 metal routing layers and one top metal layer. The die B to die E have additional back metal layer which are used to connect different dies. A tungsten layer is also added to metal layers to simulate the connection between metals. The total thickness of the structure is 284.3µm.

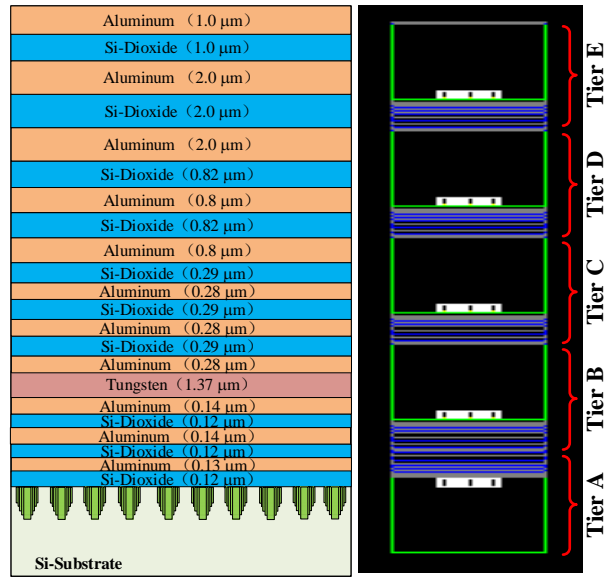

Fig.5 The simulation models of 3D SRAMs (not all SVs are depicted in this figure due to the limited content).

### 3.2 Physics Process and Heavy Ion Sources

In this work, the Monte Carlo simulations employ program package based on Geant4 which is a complete simulation toolkit for emulation the passage of particles through matter [17]. To reflect the actual physical process between incident ions and device, in the Monte Carlo simulations, ionization, nuclear elastic and inelastic reactions, and screened Coulomb scattering were involved in the physical process [18]. For example, G4EmStandardPhysics class which can simulate the electro-magnetic processes accurately are adopted in physics list. G4ionIonisation and G4hIonisation are used to simulate the continuous energy loss due to ionization and the delta rays produced by charged ions. G4HadronElasticeProcess and G4HadronInelasticProcess are used to simulate the process of nuclear reaction. G4ScreenedNuclearRecoil are used for sensitive volumes to calculate the non-ionizing energy loss during atomic motion. G4hMultipleScattering are adopted to simulate the elastic scattering of charged particles and electrons.

The heavy ion species and energy are referred the one used at Brookhaven National Laboratory (BNL) and Heavy Ions Research Facility in Lanzhou (HIRFL). The parameters of heavy ion beams, including ion species, their associated energies, effective LET and stopping ranges in silicon are showed in Table1. All those parameters are generated using SRIM2013 [19]. During each Geant4 simulation, total number of $10^5$ mono-energetic ions randomly strike the surface of 3DIC model at normal incidence for each ion type and energy.

Table 1 Heavy ion irradiation parameters in silicon used during Monte Carlo simulations

| Ions | Energy(MeV/u) | Effective LET (MeV·cm$^2$·mg$^{-1}$) | Range(μm) |
|---|---|---|---|
| $^{39}$Ar | 96.15 | 2.051 | 4570 |
|  | 35.90 | 4.269 | 867.28 |
|  | 16.67 | 7.208 | 260.16 |
| $^{132}$Xe | 106.06 | 15.63 | 2390 |
|  | 68.18 | 20.95 | 1190 |
|  | 34.09 | 32.62 | 423.34 |
| $^{209}$Bi | 71.77 | 42.19 | 1080 |
|  | 38.28 | 58.57 | 464.63 |

### 3.2 Procedure of SEUs calculation

Fig.6 shows the calculation procedure of SEU cross sections for each die. An incident ion is projected from particle gun and strikes the surface of simulation model. The collected total energy in SVs can be calculated as follows:

$$E_{Sens,NMOS} = \sum_{i=1}^{m} \alpha_i E_{SV_i,NMOS} \tag{3}$$

$$E_{Sens,PMOS} = \sum_{i=1}^{m} \alpha_i E_{SV_i,PMOS} \tag{4}$$

where m is the number of nested SV, $E_{SV_i,NMOS}$ and $E_{SV_i,PMOS}$ is the collected energy of $i$th SV for NMOS and PMOS transistor, respectively. $E_{Sens,NMOS}$ and $E_{Sens,PMOS}$ are the total collected energy for the sensitive NMOS and PMOS during one ion striking. The deposited energy of sensitive transistors in each SRAM due to the interaction between ion and devices are converted to charge by using Eg. (5) and Eg. (6). Where, the $Q_{collected,NMOS}$ and $Q_{collected,PMOS}$ are the collected charges for NMOS and PMOS, respectively. The results are recorded to ROOT for integrated counts.

$$Q_{collected,NMOS} = (1/22.5 MeV) \times E_{Sens,NMOS} \tag{5}$$

$$Q_{collected,PMOS} = (1/22.5 MeV) \times E_{Sens,PMOS} \tag{6}$$

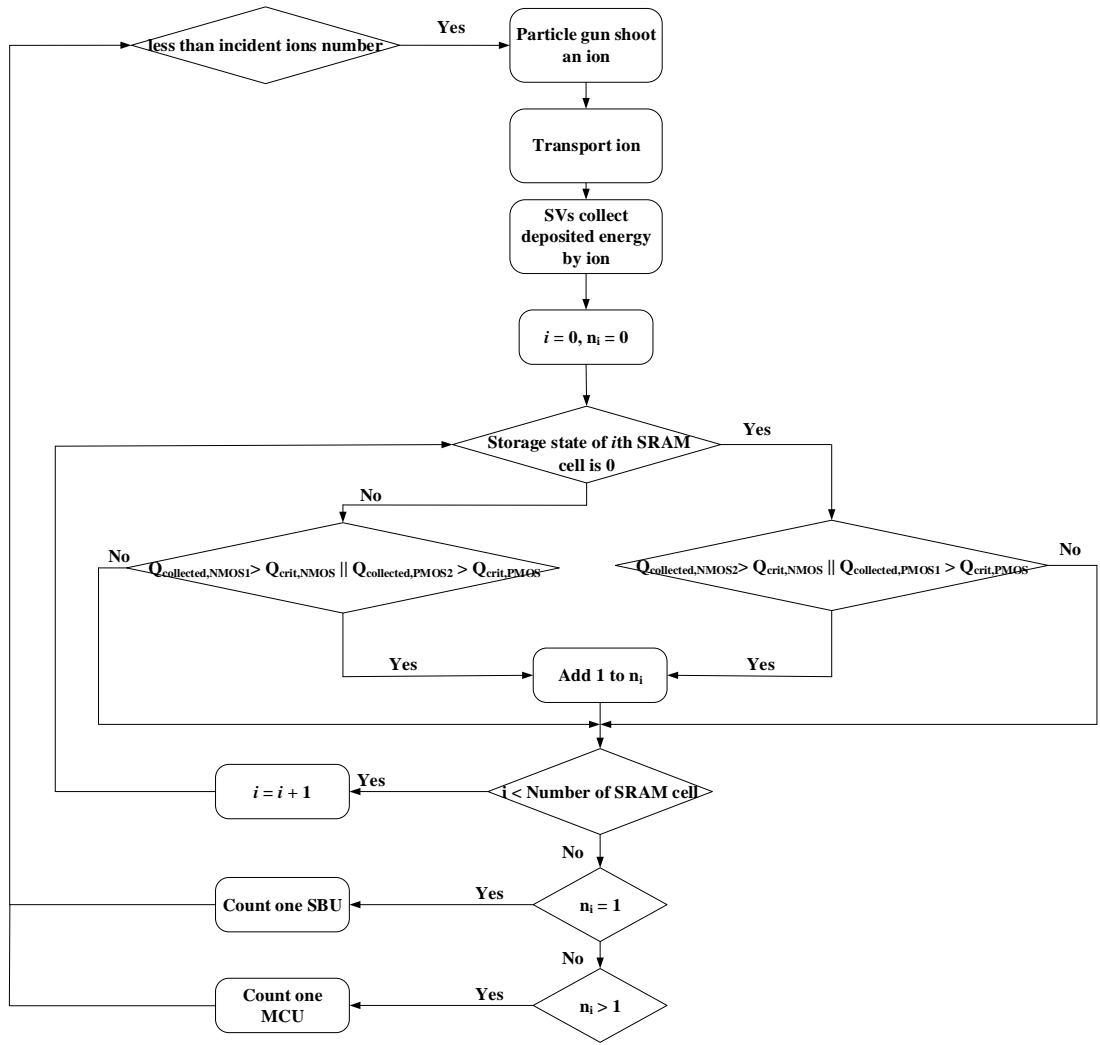

Fig. 6 The procedure of SEUs calculation in Monte Carlo simulation.

Critical charge ($Q_{crit}$) is defined the minimum charge to generate one bit upset for a SRAM. Usually the structure of device, doping concentration and reverse bias voltage govern the amount of critical charge. The critical charge used in this work is obtained by combining SPICE and TCAD simulation. The sensitive transistor models of the SRAM cell are built using TCAD. The LET of ions striking the center of drain of the sensitive transistors gradually increases, until a node state have changed from 0 to1 for NMOS or from 1 to 0 for PMOS. The LET that cause the storage state to change are considered the threshold LET, and the charge quantity collected by drain of the sensitive transistor is the critical charge. For this work, when LET of incident ions for NMOS reach to 0.26 MeV·cm$^2$/mg, the storage state begins to change. The amount of collected charges in this time is 1.5fC, same process work on the PMOS transistor, the critical charge is about 4.2fC.

When the collected charges resulting from deposited energy have exceeded critical charge, an SEU are induced, as show in Eg.(7) and Eg.(8). A SEU event, including single bit upset or multi-cell upset, at the $i$th incident particle is denoted by $n_i$. When $i$th incident ion has penetrated SVs region, either the collected charges of the sensitive transistor NMOS1 or PMOS2 for different storage state in node Q for the SRAM which shown in Fig. 1 exceed $Q_{crit,PMOS}$ or $Q_{crit,NMOS}$, one-bit flip are supposed to be induced. In this work, we suppose the proportion of storage state in node Q is same for 0 and 1 for simplifying the analysis process. After every particle event, all SVs that located in each die are executed this procedure. The number of SEUs and MCUs are counted in this process.

$$n_i = \begin{cases} 0 & Q_{collected,NMOS2} < Q_{crit,NMOS} \,\&\, Q_{collected,PMOS1} < Q_{crit,PMOS} \, (i=1,\cdots,N), \\ 1 & Q_{collected,NMOS2} \geq Q_{crit,NMOS} \,||\, Q_{collected,PMOS1} \geq Q_{crit,PMOS} \, (i=1,\cdots,N) \end{cases} \quad \text{for state 0} \quad (7)$$

$$n_i = \begin{cases} 0 & Q_{collected,NMOS1} < Q_{crit,NMOS} \,\&\, Q_{collected,PMOS2} < Q_{crit,PMOS} \, (i=1,\cdots,N), \\ 1 & Q_{collected,NMOS1} \geq Q_{crit,NMOS} \,||\, Q_{collected,PMOS2} \geq Q_{crit,PMOS} \, (i=1,\cdots,N) \end{cases} \quad \text{for state 1} \quad (8)$$

Consequently, SEU cross sections for each die can be given by

$$\sigma_{tierj}(Q_{crit}) = \sum_{i=1}^{N} n_i / (F \times N_b \times \cos(\theta)) \quad (9)$$

Where $F$ is the beam fluence with the unit of ions/cm$^2$, $N_b$ is the device capacity and $\theta$ represents the incident angle from the normal to the surface plane of the device, $\theta=0$ in our work due to the normally incident ions.

## 4. Simulation results and discussion

### 4.1 Heavy ion upset cross section

The upset cross sections of five dies for 3D SRAMs are presented in Fig.7. It demonstrates that obvious difference in the upset cross sections of five dies at low LET. The difference reach to two orders of magnitude at LET=4.269MeV·cm$^2$/mg. This maximum difference rise to four orders of magnitude when LET come down to 2.051 MeV·cm$^2$/mg. However, for heavy ions LET arriving at 15.63MeV·cm$^2$/mg, the differences nearly disappear in high LET for die B to die E, however the die A shows a smaller upset cross section compare with other dies.

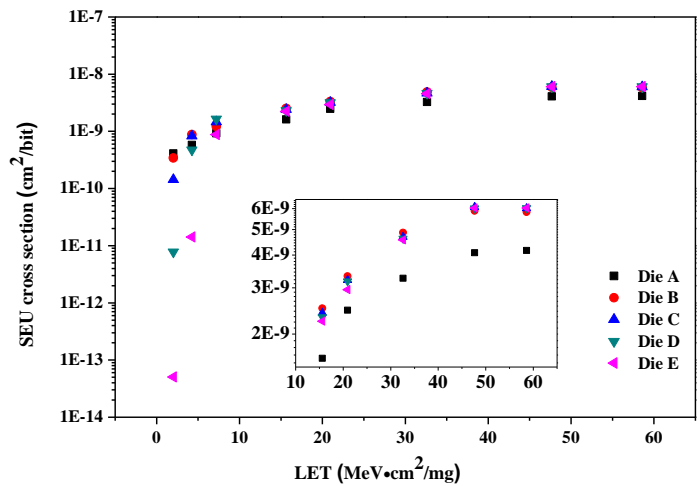

Fig. 7. SEU cross sections versus LET for 3D SRAMs. The Die A to Die E represent the upset cross sections of each die for 3D SRAMs.

In order to explain the phenomenon, Fig. 8 shows the relationship of integrated counts with deposited charge for Ar (LET=4.269 MeV·cm$^2$/mg) and Xe (LET=20.95 MeV·cm$^2$/mg), respectively. When LET is 4.269 MeV·cm$^2$/mg, the high-energy tails of the integrated counts in Fig.8 (a) presented significant discrepancies. That means the sensitivity of each die show obvious difference at low LET. And for LET reach to 20.95 MeV·cm$^2$/mg, the high-energy tails of the integrated counts begin converge together. The sensitivity of each die for 3D SRAMs tends to be similar. The cross sections of each die are getting to be close. However, the curve of integrated counts of die A shifts to left, which indicates the sensitivity of die A decrease comparing with other dies. That explains the change trend of the upset cross section of die A.

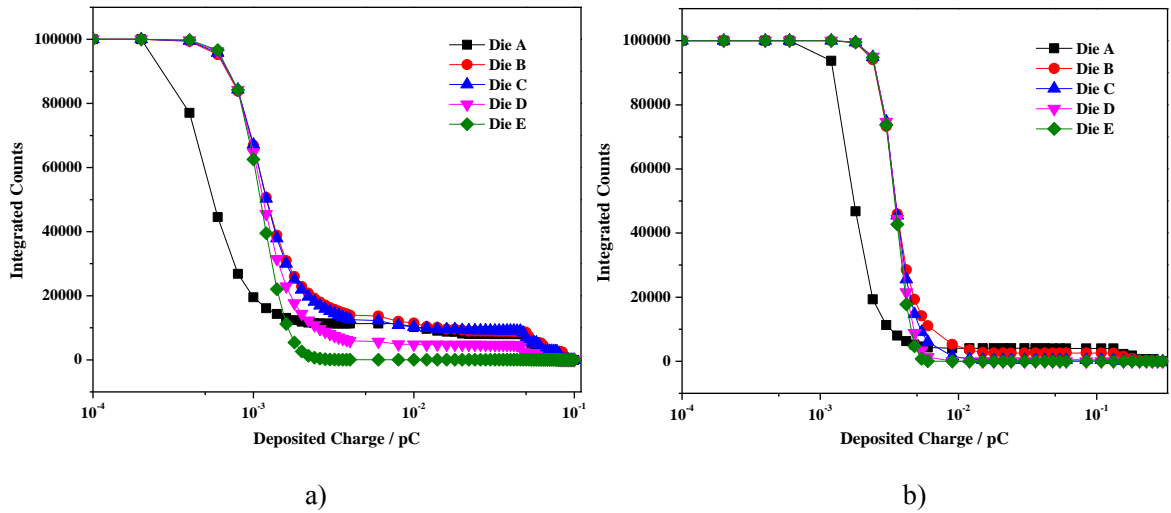

Fig 8. Integrated counts spectrum of deposited charge for the (a) Ar (4.269 MeV•cm$^2$/mg), (b) Xe (20.95 MeV•cm$^2$/mg).

### 4.2 MCUs

Fig.9 indicates that the percentage of SBUs and MCUs for Bi (71.77MeV/u and 38.28MeV/u). The simulation results show two phenomena which are concerning. The first one is that the percentage of MCUs increases as the increasing LET. The average percentage of MCUs of five dies increases from 17.2% to 32.95% when LET increases from 42.19 MeV·cm$^2$/mg to 58.57 MeV·cm$^2$/mg. The proportion has almost doubled. The occurrence of MCUs is mainly caused by nuclear reactions that generate secondary particles and delta rays. The secondary particles can be further ionized. Meanwhile delta rays can release more electrons with higher LET striking ions. Those electrons deposit their energy in more wide SVs and induced the collection of surrounding SVs, thus result in MCUs.

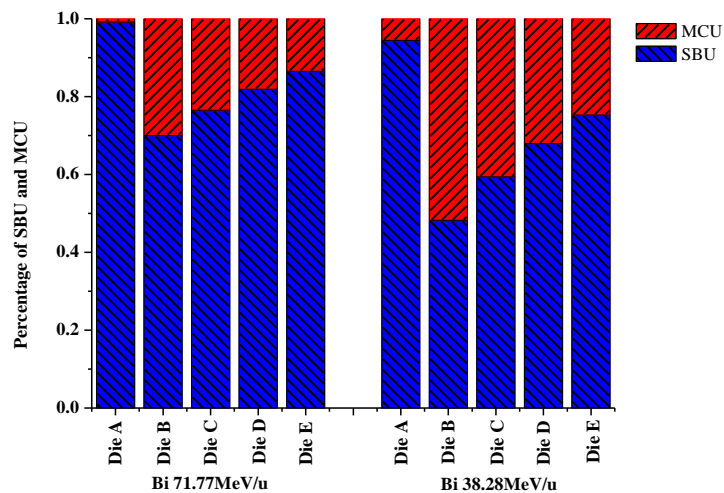

Fig.9. Percentage of SBUs and MCUs of five dies for 3D SRAM versus different energy Bi ions.

The simulation results also indicate that there are great differences in percentage of SBUs and MCUs for face-to-face and back-to-face structure. For Bi (LET=58.57 MeV•cm$^2$/mg), the average percentage of MCUs in back-to-face structure is 37.3%, but in face-to-face structure the ration is only 5.54%. For die E to die B, which are bonded by back-to-face, the percentage of MCUs increases significantly due to incident heavy ions with a more strongly scattering in the high-Z material. Besides this, the speed of Bi in die B is smaller than that in die E, the low speed increases the time of particle penetrating the SVs and raise the probability of interaction between Bi ion and tungsten. That makes the Bi scatter more diffusely. However, the percentage of MCUs in die A is very low. When ions cross over the face-to-face structure, the deposited energy in die A decrease so much that collected charges by adjacent SVs hardly exceed the critical charge, this phenomenon also can induce from results of Fig.8. The integrated counts curve come to left prove that the tolerance of die A is increasing. The results also convince that the impact of different die on sensitivity of MCUs is greater than that of SBUs.

**4.3 Comparison of planar and 3D SRAM**

In order to investigate the different sensitivity of SRAMs integrated with planar process technology and 3D technology, the simulation model of planar SRAMs is built. The planar SRAMs structure is constructed by removing interlayer dielectric isolation, metal interconnect layer and substrate of die B to die E in 3D structure which shown in Fig. 5. Aforementioned Monte Carlo method and procedure of SEUs calculation during 3D SRAMs simulations are used in this planar SRAMs simulation. The upset cross sections of both planar process and experiment as a function of LET are shown in Fig.10 a). The experimental data are from literature [20] which presented the heavy ions experimental results on SRAMs processed with 45nm bulk technology. The comparison results indicate that upset cross sections of planar process can be accurate to the experiment data, thus the building method of our models can effectively estimate the sensitivity of SRAMs. Fig.10 b) presented that the comparing results of upset cross sections of planar and 3D SRAM. Results demonstrate that the same sensitivity of planar process and 3D integrated process for high LET. For low LET, the upset cross sections of planar SRAMs are consistent with that of die A. Those results indicate that the vulnerability of 3D integrated process is not worse than that of planar process. However, for low LET, the sensitivity of each die presented obvious discrepancies, thus should give a full consideration in designing IC using 3D process due to more complex situation in low LET.

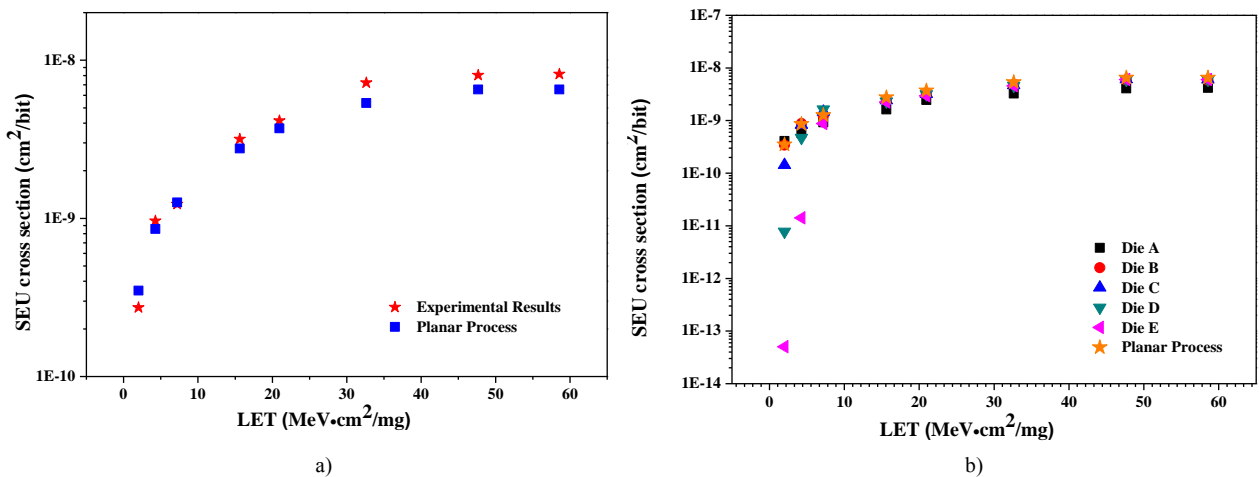

Fig. 10. SEU cross sections versus LET for 3D SRAM, planar and experimental data. Die A to die E represent the upset cross sections of each die of 3D SRAMs, planar process represents the upset cross sections of planar SRAMs.

# 5. Conclusion

Heavy ions induced single event effects have researched extensively and adequately for planar technology. However, the studies about impact of heavy ions on 3D integrated circuit technology are deficient. However, it is indispensable to evaluate the sensitivity of 3DIC due to its potential application in aerospace and military domain. In this paper, heavy ions induced single event upset sensitivity of 3D SRAMs are investigated. The SEU cross sections for heavy ions with different species, energies and LETs are simulated using Monte Carlo methods. The simulation results show that the upset cross sections of 3D SRAMs are approximate for each die in high LET, but obvious differences were observed in low LET. The percentage of MCUs also presented different trend in face-to-face and back-to-face structures. The comparison of simulation results of 3D SRAMs with planar SRAMs and experiment data show that the sensibility of 3D SRAMs is almost same to the planar SRAMs. 3D IC technology is expected to be an important application in the aerospace and military field. However, more attentions need to give the investigation of sensitivity of 3DIC due to differences of upset cross sections in low LET. In addition, the differences of percentage of MCUs for face-to-face and back-to-face structures are needed to further research.